\documentstyle[12pt]{article}
\begin{document}
\begin{flushright}
{ITEP-TH-54/96}
\end{flushright}
\vspace*{5mm}
\begin{center}
{\large \bf FUNDAMENTAL ROLE OF EXPERIMENTS IN HIGH ENERGY PHYSICS}

\vspace{3mm}

{\large L.B.Okun}\\
State Research Center\\
"Institute of Theoretical and Experimental
Physics".\\
Moscow. 117218. Russia

\vspace{5mm}

{\bf Abstract}

\end{center}

\vspace{2mm}

An introductory talk at a Meeting on results and prospects of
collaboration of Russian research organizations with European
organization for nuclear research (CERN), which took place at the
Ministry of science and technical policy of Russia on 8 July, 1996.

\vspace{2cm}

{\Large\bf Introduction.}

\vspace{5mm}

The purpose of this talk is to give a general review of High Energy
Physics, of its basic notions and terminology. The terminological
barrier is a serious obstacle in understanding any science. I will be
unable to destroy it, but I hope, that I will help you, at least
partly, to overcome it. The talk is organized as answers to 12
questions:

\begin{enumerate}
\item When was the High Energy Physics  born?
\item Why do we need the highest energies?
\item Which particles are called fundamental (elementary)?
\item What is the difference between bosons and fermions?
\item How  does the "Mendeleev Table" of fundamental particles look
like?
\item Why are the fundamental bosons needed?
\item Why is the first generation of fermions needed?
\item Why are the second and the third generations needed?
\item What is a collider? What has been discovered with colliders?
\item What do we expect from the future colliders?
\item Experiments not at the highest energies -- are they needed?
\item To understand the fundamental particles -- why is it necessary
for the mankind?
\end{enumerate}

\section{When was the High Energy Physics born?}

It was born twice. First time -- 100 years ago, when in 1896 the
radioactivity was discovered. Second time -- 50 years ago, after
World War II, when the first large accelerators of charged particles
started to create new elementary particles.

At the dawn of the century -- when X-rays, radioactivity and atomic
nucleus were discovered -- high energies stretched from thousands to
millions of electrovolts (from KeV's to MeV's). At present they
stretch from billions to trillions of eV's (from GeV's to TeV's). Let
me comment the notations: K -- kilo, M -- mega, G -- giga, T -- tera;
$1 \mbox{\rm KeV} = 10^3$ eV, $1 \mbox{\rm MeV} = 10^6$ eV,
$1 \mbox{\rm GeV} = 10^9$ eV, $1 \mbox{\rm TeV} = 10^{12}$ eV.
$1 \mbox{\rm eV} = 1 e \times 1 \mbox{\rm V}$ --
is the energy, which an
electron acquires by crossing in accelerating difference of
potentials of 1 Volt. Let me remind that $1 \mbox{\rm Amper} = 6
\cdot 10^{18}$ electrons/sec. Thus the values of energies under
discussion may seem to be quite non-impressive. But they are very
large, if one takes into account, that they are carried by single
particles!

\section{Why do we need the highest energies?}

In order to create and to study fundamental particles. You know, of
course, the famous formula by Einstein, which literally, shook the
world:
$$
E_0 = mc^2 \;\; ,
$$
where

~~~ $E_0$ -- is the rest energy of a body (the index zero indicates
that the body is at rest, its velocity being equal to zero),

~~~~ $m$ -- is the mass of the body,

~~~~ $c= 3 \cdot 10^8$ m/sec -- is the velocity of light.

In experiments on accelerators the kinetic energy of accelerated
particles is transformed, during collisions, into the rest energy
(mass) of created particles. The higher is the energy of an
accelerator, the heavier are the particles which it can produce.

\section{Which particles are called fundamental (elementary)?}

Those, which at present level of knowledge do not consist of more
fundamental ones ("the smallest matreshkas"). Atoms are not
elementary: they consist of electrons and nuclei. Nuclei consist of
protons $p$ and neutrons $n$. Protons and neutrons are also not
elementary, they consist of quarks of two types, $u$-quarks and
$d$-quarks:
$$
p = uud \; , \;\; n = ddu \;\; .
$$

Quarks and electrons are elementary at present level of knowledge.
Photons -- the particles of light -- are elementary as well. Quarks
and electrons belong to a family of particles called fundamental
fermions. Photons belong to another family, that of fundamental
bosons.

\section{What is the difference between bosons and fermions?}

They differ by the value of their spin. Particles are like miniature
tops. Spin is the proper rotational (angular) momentum of a particle.
Spin is measured in units of $\hbar$:
$$
\hbar = 10^{-34} \mbox{\rm Joules} \cdot \mbox{\rm sec} = 10^{-34}
\mbox{\rm kg} \cdot \mbox{\rm (m/sec)} \cdot \mbox{\rm m}
$$

In order to visualize the value of $\hbar$, imagine one gram weight,
which is fixed on a rotating 1 cm long stick so that its velocity is
1 cm/sec. And now reduce the mass by 27 orders of magnitude, the
length by 10 orders, and increase the velocity by 10 orders. (In
order to reduce a stick by 10 orders, one has to break it into two
halves, then to break, in the same way, one of the halves, and to
repeat this procedure 31 times "only".)

The constant $\hbar$ -- one of the most fundamental constants of
nature. It was introduced by German physicist Max Planck in 1990.

Bosons (named after Indian physicist Bose) have integer values of
spin in units of $\hbar$. Fermions (named after Italian physicist
Fermi) have half-integer spin. The spin of photon is $1 \hbar$, or
simply 1; the spin of electron is 1/2.

Fermions are individualists: there can exist, in a given state, only
one fermion of a given type. This property explains the pattern of
atomic levels and hence the Mendeleev Table.

Bosons are collectivists: all bosons of a given type prefer to be in
one state. This property is the basis of the laser.

The amazing properties of bosons and fermions are connected with
basic principles of the modern quantum physics. I do not know of any
simple graphic explanation of these properties.

Now we are ready to answer the key question of this talk.

\section{How does the "Mendeleev Table" of fundamental particles look
like?}

The Table contains 16 particles: 4 bosons and 12 fermions.

\newpage
\begin{center}

{\it Fundamental bosons.}

\end{center}
\vspace{2mm}

The photon, $\gamma$, the $W$ and $Z$ bosons, the gluon, $g$. All of
them have spin 1. The photon, gluon and $Z$ boson are electrically
neutral: their electric charge $Q$ is equal to zero. For $W$ bosons
$Q = \pm 1$.

\vspace{3mm}
\begin{center}

{\it Fundamental fermions.}

\end{center}
\vspace{2mm}

Twelve fermions are subdivided into three generations, two quarks and
two leptons in each. (The term leptons means electron and its
"relatives".) The first three columns of the following table
represent three generations of fundamental fermions. The fourth
column shows their electric charges $Q$.

\vspace{3mm}
\begin{center}
\begin{tabular}{|ll|c|c|c||c|}
\hline
generation & & 1st & 2nd & 3d & $Q$ \\ \hline
& upper & $u$ & $c$ & $t$ & +2/3 \\
quarks & & & & & \\
& lower & $d$ & $s$ & $b$ & -1/3 \\  \hline
 & neutrinos & $\nu_e$ & $\nu_{\mu}$ & $\nu_{\tau}$ & 0 \\
leptons & & & & & \\
& charged leptons & $e$ & $\mu$ & $\tau$ & -1 \\
\hline
\end{tabular}
\end{center}

\vspace{3mm}

As seen from the table, there are two types of quarks: "upper" and
"lower". (The symbols $u$ and $d$ stem from "up" and "down", whilst
$t$ and $b$ -- from "top" and "bottom"; $c$ and $s$ denote so called
"charmed" and "strange" quarks.) As for leptons, they are subdivided
into neutral (neutrinos) and charged ones.

Each charged fermion has its antiparticle: $\bar{u}$, $\bar{d}$,
$\bar{c}$, $\bar{s}$, $\bar{t}$, $\bar{b}$, $e^+$, $\mu^+$, $\tau^+$.
About neutrinos this is still uncertain. It is one of the important
problems -- to establish whether or not antineutrinos are identical
with neutrinos.

Let us note that 5 fundamental fermions ($c, b, t, \tau, \nu_{\tau}$)
and 3 fundamental bosons ($g, W, Z$) have been discovered after 1973.

\section{Why are the fundamental bosons needed?}

The main role of the known fundamental bosons is to serve as
mediators of forces.

The exchange of photons creates electromagnetic forces which underlie
the atomic and molecular physics, physics of solid state and plasma,
optics, acoustics, chemistry, biology.

The exchange of gluons creates strong interactions which confine
quarks inside protons, neutrons and hundreds of other particles which
are built of quarks and gluons and which are called hadrons.

The exchange of $W$ and $Z$ bosons creates weak forces.

Theorists have no doubt that gravity is produced by exchange of
gravitons, fundamental bosons with spin equal 2. However to observe
these particles is extremely difficult. Even our grand-grand children
will be unable to detect them.

\section{Why is the first generation of fermions
needed?}

The world around us and we ourselves are built from electrons and
$u$- and $d$-quarks. Without weak interactions with participation of
$\nu_e$ there would be no stars, no sun, no life. A complex chain of
nuclear reactions involving electron neutrinos inside the sun
transforms hydrogen into helium and then into heavy elements. Protons
"burn" with release of energy and neutrinos:
$$
2 e^- + 4p \to ~^4He + 2\nu_e + 27 \; \mbox{\rm MeV} \;\; .
$$
The flux of solar neutrinos is enormous: 70 billion per second per
each cm$^2$ on the earth. But we are practically transparent for
them. Only special multikilotonne detectors can capture a few
particles from this flux. (One of such detectors operates in
our country, in the Baksan valley.)
The number of captured $\nu_e$'s turned out
to be approximately a factor of two less, than  had been expected.
Maybe, on their way from the center of the sun to the earth,
$\nu_e$'s partly transform into $\nu_{\mu}$'s or $\nu_{\tau}$'s,
which leave no trace in detectors, which are sensitive only to
$\nu_e$'s?

\section{Why are the second and the third generations of fermions
needed?}

We have no definite answer to this question. Maybe, they are needed
in order to preserve some amount of electrons and protons from the
time of the "big bang". Otherwise the world would consist only of
photons and neutrinos. (On an average per one proton, there is one
electron, one billion of photons, with energy $3\cdot 10^{-4}$ eV in
the form of radiowaves, and approximately the same number of
neutrinos.)

\section{What is a collider? What has been discovered by using
colliders?}

Collider is a machine for accelerating, storing (not always) and
head-on colliding two beams of particles. (In an ordinary
accelerator, not a collider, there is only one beam, which hits a
fixed target.) In the head-on collisions in colliders the kinetic
energy is transformed into the rest energy of created particles in
the most effective way.

The particles discovered with colliders are: $t$- and partly
$c$-quark, $\tau$-lepton, gluon, $W$ and $Z$-bosons.

The masses of $W$ and $Z$ bosons are 80 GeV and 91 GeV, respectively.
These bosons have been discovered at CERN on a specially built for
this purpose proton-antiproton collider, with energies of particles
in each beam 270 GeV.

20 million $Z$ bosons have been created and
detected at CERN, in 1989--1995, on the LEP I collider. In a circular
tunnel of LEP I, with 27 km circumference, beams of electrons and
positrons with energies 45.5 GeV collided head-on. Their energy was
fully used to create $Z$ bosons.

In 1994 at the proton-antiproton collider Tevatron (USA), the
heaviest of known particles have been created -- $t$-quark, its mass
being about 175 GeV. The energy of particles in each of the two beams
of Tevatron is about one TeV, hence the name of the collider. May I
remind you that the mass of a proton is 0.94 GeV, while that of
electron is 0.5 MeV.

\section{What do we expect from the future colliders?}

In the same ring, where LEP I was operating, a new collider started
to run in the fall of 1999. The energies of $e^-$ and $e^+$ will
reach in it 96 GeV. In year 2000, in the same tunnel the construction
of a new machine -- Large Hadron Collider (LHC) -- will start. The
colliding particles in this machine will be protons with energy 7
TeV.

First of all, physicists expect, that  with the new colliders, a
new particle called Higgs boson (or simply, higgs) will be
discovered. (P.Higgs is a contemporary British theorist.) The spin of
the higgs must be equal to zero. Its mass cannot be predicted in a
definite way. Most probable, however, that higgs is heavier than $W$
boson, but lighter than the top-quark. The higgs plays a central role
in the modern theory of matter. According to the theory, all
fundamental particles acquire their masses through their interaction
with the higgs. The discovery of the higgs would allow physicists
to come closer to understanding the nature of mass.

Another promising direction is supersymmetry, according to which to
each of the known particles there correspond a "superpartner": a
particle with spin differing by 1/2. Thus superpartner of a fermion
is a boson, whilst superpartner of a boson is a fermion. The
supersymmetry is strongly broken in the nature. It is expected, that
the masses of "superpartners" of the known particles lie mainly in
the interval from 100 GeV to 1 TeV.

Third direction is the possible compositness of our fundamental
particles, which may reveal itself  at higher energies.

Finally, one has not to forget about "expected surprises". In the
past, many important phenomena were unexpectedly discovered at
accelerators, which had been built for other purposes and did not
have these phenomena on their "to be discovered" lists.

\section{Experiments not at the highest energies -- are they needed?}

Yes! The idea that all efforts should be concentrated on the highest
energy colliders is an erroneous one.

The colliders represent the direction of the principle attack, but we
are fighting simultaneously on several fronts. The answers to many
crucial questions cannot be, in principle, obtained with colliders.
They might be obtained either in non-accelerator experiments, or in
experiments on low energy fixed target accelerators. (In the latter
case they may be considered as "pockets of resistance".) Here are a
few examples:
\begin{itemize}
\item Search for neutrino masses.
\item Study of solar neutrinos (the neutrino monitoring of the sun
could be exceptionally important from practical point of view).
\item Search for proton decay.
\item Search for transformation of neutrons into antineutrons in
vacuum.
\item Study of the asymmetry between particles and antiparticles.
\item Study of light hadrons in order to understand the mechanism of
confinement of quarks in hadrons.
\end{itemize}

It should be stressed that for the study of light hadrons the
existing accelerators in Russia could be highly effective.

\section{To understand the fundamental particles -- why is it necessary
for the mankind?}

In order to ascertain the basic principles of nature (as an ideal
-- one principle, from which follow all fundamental laws).

In order to understand the birth of the universe and its future. (One
TeV corresponds to the temperature of the universe at the age of one
picosecond).

High Energy Physics is a tuning fork of the intellectual sphere of
mankind.

There exists at present a unique community of engineers, experimental
and theoretical physicists, which can solve these problems. It should
not be lost!

\end{document}